\documentclass[reprint,superscriptaddress,amsmath,amssymb,aps,pra,floatfix]{revtex4-1}

\usepackage{graphicx}
\usepackage{dcolumn}
\usepackage{color}
\usepackage{bm}
\usepackage{hyperref}
\hypersetup{colorlinks= true,linkcolor=blue,citecolor=blue} 
\usepackage{verbatim}

\begin{document}

\newcommand{\bblue}{\color{black}}

\title{\bblue Thermal properties of NbN single-photon detectors}
\author{E.M.~Baeva}
\affiliation{National Research University Higher School of Economics, 20 Myasnitskaya St, Moscow, 101000, Russia}
\affiliation{Moscow State University of Education, 29 Malaya Pirogovskaya St, Moscow, 119435, Russia}
\author{M.V.~Sidorova}
\affiliation{Moscow State University of Education, 29 Malaya Pirogovskaya St, Moscow, 119435, Russia}
\affiliation{DLR Institute of Optical Systems, Rutherfordstrasse 2, Berlin, 12489, Germany}
\author{A.A.~Korneev}
\affiliation{National Research University Higher School of Economics, 20 Myasnitskaya St, Moscow, 101000, Russia}
\affiliation{Moscow State University of Education, 29 Malaya Pirogovskaya St, Moscow, 119435, Russia}
\author{K.V.~Smirnov}
\affiliation{National Research University Higher School of Economics, 20 Myasnitskaya St, Moscow, 101000, Russia}
\affiliation{Moscow State University of Education, 29 Malaya Pirogovskaya St, Moscow, 119435, Russia}
\affiliation{LLC Superconducting nanotechnology (Scontel) , 5/22 Rossolimo St,Moscow, 119021, Russia}
\author{A.V.~Divochy}
\affiliation{Moscow State University of Education, 29 Malaya Pirogovskaya St, Moscow, 119435, Russia}
\affiliation{LLC Superconducting nanotechnology (Scontel) , 5/22 Rossolimo St,Moscow, 119021, Russia}
\author{P.V.~Morozov}
\affiliation{LLC Superconducting nanotechnology (Scontel) , 5/22 Rossolimo St,Moscow, 119021, Russia}
\author{P.I.~Zolotov}
\affiliation{National Research University Higher School of Economics, 20 Myasnitskaya St, Moscow, 101000, Russia}
\affiliation{Moscow State University of Education, 29 Malaya Pirogovskaya St, Moscow, 119435, Russia}
\affiliation{LLC Superconducting nanotechnology (Scontel) , 5/22 Rossolimo St,Moscow, 119021, Russia}
\author{Yu.B.~Vakhtomin}
\affiliation{Moscow State University of Education, 29 Malaya Pirogovskaya St, Moscow, 119435, Russia}
\affiliation{LLC Superconducting nanotechnology (Scontel) , 5/22 Rossolimo St,Moscow, 119021, Russia}
\author{A.V.~Semenov}
\affiliation{Moscow State University of Education, 29 Malaya Pirogovskaya St, Moscow, 119435, Russia}
\author{T.M.~Klapwijk}
\affiliation{Moscow State University of Education, 29 Malaya Pirogovskaya St, Moscow, 119435, Russia}
\affiliation{Kavli Institute of Nanoscience, Delft University of Technology, Delft 2628 CJ, The Netherlands}
\author{V.S.~Khrapai}
\affiliation{National Research University Higher School of Economics, 20 Myasnitskaya St, Moscow, 101000, Russia}
\affiliation{Moscow State University of Education, 29 Malaya Pirogovskaya St, Moscow, 119435, Russia}
\author{G.N.~Goltsman}
\affiliation{National Research University Higher School of Economics, 20 Myasnitskaya St, Moscow, 101000, Russia}
\affiliation{Moscow State University of Education, 29 Malaya Pirogovskaya St, Moscow, 119435, Russia}

\begin{abstract}
{\bblue We investigate thermal properties of a NbN single-photon detector capable of unit internal detection efficiency. 
Using an independent calibration of the coupling losses we determine the absolute optical power absorbed by the NbN film and, via a resistive superconductor thermometry, the thermal resistance $Z(T)$ of the NbN film in dependence of temperature. In principle, this approach permits a simultaneous measurement of the electron-phonon and phonon-escape contributions to the energy relaxation, which in our case is ambiguous for their similar temperature dependencies. We analyze the $Z(T)$ within the two-temperature model and impose an upper bound on the ratio of electron and phonon heat capacities in NbN, which is surprisingly close to a recent theoretical lower bound for the same quantity in similar devices. }
\end{abstract}

\maketitle

\section{\label{sec:Introduction}Introduction}

{\bblue State of the art superconducting single-photon detectors (SSPDs) comprise a sensitivity to infrared photons, a near-100\% detection efficiency~\cite{Marsili2013,Kahl2015,Smirnov2018} and a counting rate of hundreds of MHz~\cite{Pearlman2005}, making them suitable for a wide range of applications \cite{natarajan}. To combine all these high performances in one device, however, is a challenge and requires not only technical advancements but a clear understanding of the physics behind the single-photon detection, see for a recent review~\cite{bartolf2015fluctuation}. 

The problem of SSPD theory lies in a complex interplay of resistive and thermal properties of a thin superconducting film under strong nonequilibrium~\cite{Klapwijk2017}. Understanding the resistive properties has two major complications. First, the superconducting films usually used in SSPDs are strongly disordered, with a sheet resistance in the range approaching a superconductor-insulator transition~\cite{Gantmakher2010}. Second, the physics of the resistive transition in thin superconducting films is closely related to the Berezinskii-Kosterlitz-Thouless (BKT) theory \cite{Beasley1979, Halperin1979}, which connects the temperature dependence of the resistance with the density of free topological excitations --- superconducting vortices. The thermal aspect is yet more complex, for the conversion of the absorbed photon energy in a current biased film interconnects the electronic excitations, the Cooper pair condensate with vortices and the phonon system.

The very beginning of the detection event manifests in a highly non-equilibrium region within the superconducting film, routinely called the hot-spot (HS). In the simplest interpretation of a so-called ``geometric HS model"~\cite{sspd}, the role of the HS is to create a normal conducting region, which causes a redistribution of the current density incompatible with the zero resistance state. Recent theoretical work~\cite{Zotova2014,vodolazov} considers the situation as much less trivial owing to a role played by the current-crowding effect and the superconducting vortices, accordingly dubbed in Ref.~\cite{korneeva2018} as the ``photon-generated superconducting vortex model". In this  model, the emerging HS captures a single vortex or a vortex--antivortex pair, depending on the position of the HS across the superconducting strip. Above a certain threshold bias current, the vortices can unbind from the HS and pass across the strip in an avalanche manner, causing the dissipation and expansion of the HS. The detection efficiency of an SSPD depends on how the initial photon energy is redistributed between the electron and phonon systems throughout the HS evolution. The ``photon-generated superconducting vortex model"~\cite{vodolazov} relates the efficiency with the ratio of heat capacities of electron and phonon systems in a universal way, which emphasizes the role of thermal properties of a superconductor used in SSPD. }

{\bblue In this work, we analyze a NbN nanowire-based SSPD capable of unit internal detection efficiency in terms of its thermal response under continuous irradiation with the infrared light. The key ingredient of our approach is the  knowledge of the absolute amount of energy absorbed by the film, which is achieved via an independent calibration of the apparatus' coupling losses. Using resistive superconductor thermometry in zero and finite magnetic fields, we obtain the thermal resistance $Z(T)$ of a NbN film and analyze it within a two-temperature model. Potentially, our approach allows to extract the individual contributions of the electron-phonon relaxation and of the phonon escape into substrate to $Z(T)$. This turns out difficult in present experiment, for close temperature dependencies of the two contributions. Instead, we impose an experimental upper bound on the ratio of electron and phonon heat capacities, which is surprisingly consistent with the lower bound implied by the model calculations for similar devices~\cite{vodolazov}.}

\section{\label{sec:Methods}Methods and experimental details}

\subsection{\label{sec:Theory}Two-temperature model}

\begin{figure}[h!]
    \centering
    \includegraphics[width = 0.8\columnwidth]{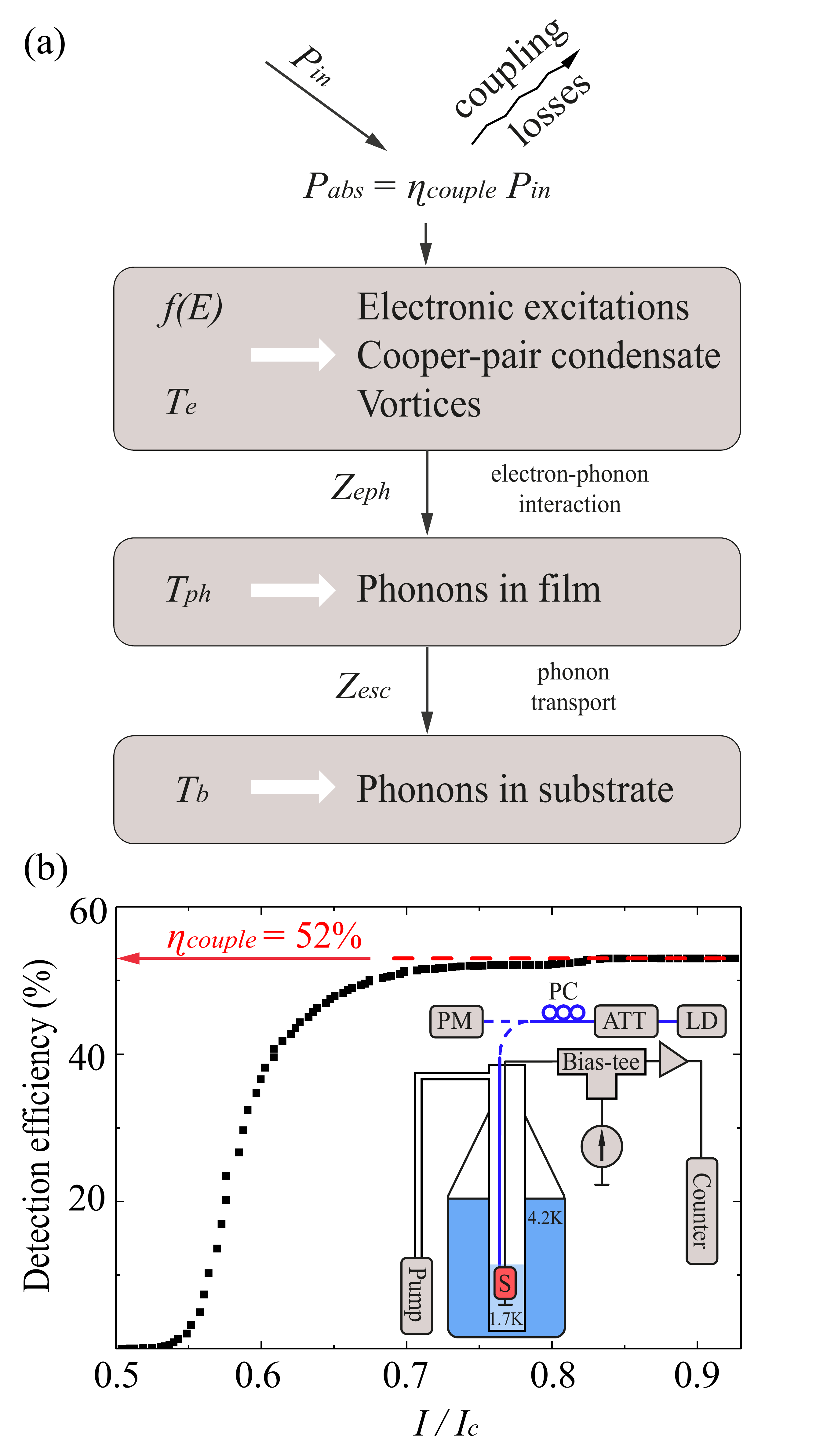}
    \caption{(a): Two-temperature model of a metal-superconducting film on an insulating substrate. Only a fraction of the incident optical power $P_{in}$ is absorbed by the film as determined by the coupling efficiency of the apparatus $P_{abs}=\eta_{couple}P_{in}$. The electron subsystem under illumination can be described by the {\bblue equilibrium} Fermi-Dirac distribution with an effective electron temperature $T_e$ exceeding the phonon  temperature in NbN $T_{ph}$ and the bath $T_b$ temperature. Further, it is assumed that the $T_e$ is the only parameter that controls the resistivity of the superconducting film. $Z_{eph}$ and $\bblue Z_{esc}$ are the thermal resistances, which determine the heat exchange of the electrons with the acoustic phonons in NbN and of the phonons in NbN with the substrate, respectively. (b): Measured SSPD detection efficiency $D_E$ at $1.55\,\mu$m wavelength at $T_b=1.7$\,K vs the bias current in units of the critical current. The saturation of the bias current dependence signals the regime of unit internal efficiency, which enables us to calibrate the  coupling losses as $D_E=\eta_{couple}$. \textcolor{black}{Inset: The sketch of the experimental setup. S denotes the sample, LD -- the laser diode, PC -- the polarization controller, PM -- the power meter and ATT -- the attenuator.}}
    \label{fig: thermalresistors}
\end{figure}

{\bblue As discussed above, the resistive response of an SSPD is based on a complex nonequilibrium process of conversion of the absorbed photon energy into the energy of the Cooper pair condensate with superconducting vortices, the electronic excitations and the phonon system. Focusing here on the thermal properties in the vicinity of the resistive transition, we use} a more reduced approach and consider the system in terms of the commonly used two-temperature model (\autoref{fig: thermalresistors}). {\bblue With certain reservations (see section~\ref{resistivity} for the details), one can neglect} the spatially inhomogeneous nature of the photon absorption and assume a uniform{\bblue, yet different, temperatures for the electrons, $T_e$, and for the (acoustic) phonons, $T_{ph}$.}  The radiation energy is absorbed by the electrons and transfered to the {\bblue substrate within} a time called the energy relaxation time{\bblue, that can be measured via amplitude-modulated absorption of the THz radiation (AMAR) \cite{kardakovaPRB}.} This method has been first introduced by Gershenzon \textit{et al} \cite{gershenzon1990Nb} {\bblue and since then applied to various superconducting materials with different $T$-dependences of the relaxation time, for instance} $T^{-2}$ in Nb \cite{gershenzon1990Nb} and in boron-doped diamond \cite{kardakovaPRB}, $T^{-1.6}$ in NbN \cite{gousev,gousev1994} and $T^{-3}$ in TiN \cite{kardakova2013tin}, in NbC \cite{il1998nbc} and WSi~\cite{Sidorova2018} films. 

{\bblue Similar to the AMAR approach, here we also maintain the sample at a bath temperature  $\bblue T_b$ close to the resistive transition and bias it with a DC current $I$, in order to measure the change of the resistance in response to the applied radiation, which is referred to as resistive superconductor thermometry above. We assume that the current is small enough, such that } there is a one-to-one {\bblue thermodynamic} correspondence between an increase in {\bblue $T_e$} and the film resistance. In the BKT scenario, the resistance is determined by the number of free superconducting vortices and anti-vortices, which are equally populated in a zero magnetic field and preferably polarized in one direction in a finite magnetic field~\cite{Ryzhov2017}. We, therefore, assume that the change in the vortex density caused by both the DC current and the absorbed optical power is negligible compared to its equilibrium value, which is justified in the vicinity of the resistive transition, where the topological excitations are macroscopically populated. {An important difference from the AMAR case is that in present experiment the modulation frequency of the radiation is much smaller than the intrinsic relaxation rate in NbN SSPDs, since we are interested in a time-averaged and device-averaged resistive response, which is a measure of the thermal resistance between the electron system in NbN and the bulk substrate.}

In the experiment we exploit the resistive properties of the superconductor as a function of $\bblue T_e$. {\bblue In the vicinity of the} resistive transition, where phase-coherence breaks down, we assume that the resistivity of the superconductor is {\bblue a direct measure of $T_e$. Within the  two-temperature model, the} time-dependent temperatures of the electrons $T_e$ and phonons $T_{ph}$ are obtained from the following coupled linear heat-balance equations \cite{vanneste, perrin}:
\begin{equation} 
\label{eq:1}
C_e\frac{ dT_e}{dt}=P_{abs}+P_{DC}-\frac{C_e}{\tau_{eph}}(T_e-T_{ph})
\end{equation}
\begin{equation}
\label{eq:2}
C_{ph}\frac{ dT_{ph}}{dt}=\frac{C_{ph}}{\tau_{phe}}(T_e-T_{ph})-\frac{C_{ph}}{\tau_{esc}}(T_{ph}-T_b)
\end{equation}
where {\bblue $T_b$ is the bath} temperature, $P_{DC}=IU$ is the Joule {\bblue heat} power and $P_{abs}$ is the power of the applied radiation absorbed by the film, $\tau_{eph}$ is the electron-phonon time, $\tau_{phe}$ is phonon-electron time and $\tau_{esc}$ is the phonon escape time. {\bblue Note that a usual equilibrium relation $C_{ph}/\tau_{phe}=C_{e}/\tau_{eph}$ also holds here.}

The differential equations {\bblue (\ref{eq:1}) and (\ref{eq:2})} are further simplified {\bblue under the conditions of our experiment}. {\bblue The first} is  the steady-state condition $dT_i/dt=0$, since the modulation period of the incident radiation, $1/f=1/26\,{\rm Hz} = 38$\,ms, is much greater than the typical time constants in dirty NbN, that is the energy relaxation time (10-20\,ps)~\cite{gousev1994,semenov2009} and the phonon escape time ($\sim 50$\,ps) \cite{semenov1}. {\bblue The second is the condition of small bias current, which allows us to neglect the Joule heating, $P_{DC}\ll P_{abs}$ and minimize the potential effect of the current on the creation of extra free vortices beyond the equilibrium number determined by the electron temperature  (not included in the above model). As such, the model reduces to a single equation for the total thermal resistance, $Z\equiv (T_e-T_b)/P_{abs}$:}
\begin{equation} 
\label{eq:3}
Z=\frac{\tau_{eph}}{C_e}+\frac{\tau_{esc}}{C_{ph}}=Z_{eph}+Z_{esc}
\end{equation}

Equation \eqref{eq:3} intuitively expresses the two-stage transfer of the heat from the electronic system towards the equilibrium bath in our two-temperature model, as schematically presented in Fig.~\ref{fig: thermalresistors}a. The absorbed radiation power raises the electron temperature $T_e$ with respect to the acoustic phonon temperature $T_{ph}$ in the NbN film. The difference between $T_e$ and  $T_{ph}$
is determined by the first contribution in eq.~(\ref{eq:3}), which is the electron-phonon thermal resistance in NbN  $Z_{eph}\equiv\tau_{eph}/C_e$. In turn, the finite escape time of the acoustic phonons from the NbN into the substrate causes the increase of $T_{ph}$ compared to the bath temperature  $T_b$, which results in the second contribution to the thermal resistance $Z_{esc}\equiv\tau_{esc}/C_{ph}$.

In what follows we will be interested in the ratio of the normal state thermal capacity of the electronic system, $C^n_e$, and the  thermal capacity of the phononic system in NbN $C_{ph}$. While the former quantity is straightforward related to the density of states at the Fermi level in NbN and can be obtained by all transport means as explained below, the latter is not so easily accessible. Here, we infer the lower bound for $C_{ph}$, hence the upper bound for $C^n_e/C_{ph}$, from the measured $Z$ using the eq.~(\ref{eq:3}) as follows:
\begin{equation} 
\label{eq:cecp}
\bblue
\frac{C^n_e}{C_{ph}} < C^n_e \frac{Z}{\tau_{esc}} < C^n_e \frac{Z}{\tau^{low}_{esc}},
\end{equation}
{\bblue where $\tau^{low}_{esc}$ is the lower bound for the acoustic phonon escape time in dirty NbN films adopted from the literature. Remarkably, we find that at $T=T_c$ such an upper bound is close to the recent theoretical lower bound~\cite{vodolazov} expected for the same quantity in our SSPD. This observation is one of the main results of our paper, along with the absolute measurement of $Z$.}

\subsection{\label{sec:Sample}Sample}

We perform our experiments on a $5.5\pm0.5$ nm-thick, 100 nm wide NbN nanowire patterned in a meander shape. \textcolor{black}{ The meander covers a  15$\times$15\,$\mu {\rm m}^2$ area, that roughly corresponds to the size of a light spot at the end of the optical fiber. On top of the sample the 450\,nm thick anti-reflection Al$_2$O$_3$/Si/Al$_2$O$_3$ coating is deposited. To enhance absorption of the sample we arrange an optical cavity between the NbN meander and the SiO$_2$/Si substrate, which consists of a 70\,nm thick bottom Au mirror and a 200\,nm thick dielectric Si$_3$N$_4$ layer~\cite{Smirnov2018}.} The device has a $T_c$ of 7.84 K. The sheet resistance $R_{\square}$ is measured at room temperature by using a four-point probe technique for the film before lithographic processing. The resistivity $\rho$ is derived from $\rho = R_{\square} d$, where $d$ the thickness is determined from the deposition time using a calibrated deposition rate.

The resistivity in the normal state, $\rho_n$, relevant for the diffusivity and elastic scattering in the superconducting state is determined from the resistivity above the resistive transition at $T=20$\,K.  The measured resistance ratio, \textcolor{black}{$\rho_{(300K)}/\rho_{n}=R_{(300K)}/R_{(20K)}=0.8$}, is used to convert the room temperature measurement to the actual number, which is $4.4 \cdot 10^{-6} \Omega \cdot$m. In contrast to conventional metals the resistivity at cryogenic temperatures is higher than at room temperature. The parameters of the device are summarized in \autoref{table1}.
\begin{table}[h!]
\caption{\label{table1} Parameters of the NbN film, with $d$ the thickness, $V$ the volume of the device, $T_c$ determined from the mid-point of the resistive transition, $I_c$ the superconducting critical current at T=1.7 K, $R_{\square}$ the sheet resistance at room temperature, and $\rho_n$ the resistivity at $T=20~ K$.   $D$ is diffusivity extracted from the temperature dependence of resistance at magnetic field and $N(0)$ is a single-spin density of states obtained from the transport data (\autoref{resistivity}). }
\begin{center}
\begin{ruledtabular}
\begin{tabular}{|c|c|c|c|c|c|c|c|}
$d$ & $V$& $T_c$& $I_c$ & $R_{\square}$& $\rho_n$ & $D$& $N(0)$ 
\\ nm & cm$^3$& K& $\mu$A & $\Omega / \square$& $\Omega \cdot$m  & cm$^2/$s& ${\rm eV^{-1}nm^{-3}}$ \\
\hline
 5.5 & 6.6$\cdot 10^{-13}$& 7.84& 13 &640& 4.4$\cdot 10^{-6}$ & $0.35$ & $20.3$ \\
\end{tabular}
\end{ruledtabular}
\end{center}
\end{table}

\subsection{\label{sec:Power}Absorbed power}

The {\bblue knowledge of the absolute amount of the absorbed power $P_{abs}$ is a key ingredient to determine the thermal resistance $Z$. 
In general, only a fraction of the input optical power $P_{in}$ is absorbed by the detector. In our experiment, ${\eta_{couple}=P_{abs}/P_{in}<1}$ is mainly} caused by imperfections of a multilayer \textcolor{black}{anti-reflection} coating and a misalignment of the light spot at the end of the {\bblue optical} fiber with respect to the detector area, {\bblue which is quantified as described below.}

{\bblue Even in an idealized SSPD free from all sorts of coupling losses ($\eta_{couple}=1$) the absorption of a single-photon does not necessary result in a detection event, that is a temporary switching of the device into the resistive state. In dirty superconductors with short thermalization time, including NbN and present experiment, the underlying switching mechanism is most adequately described within a hot-spot (HS) model~\cite{vodolazov}. In this case, shortly after the photon absorption the superconducting gap inside the HS is suppressed, giving rise to the nucleation of a vortex-antivortex pair inside the superconducting strip or the entrance of a single vortex from its nearest edge~\cite{Zotova2014}. Whether this promotes the detection event or not depends on the position, shape and the volume of the HS. In addition, the internal efficiency, which determines the single-photon detection probability of the idealized SSPD, should increase with the bias current $I$ owing to a current-crowding effect and heating of the superconductor by moving unbound vortices~\cite{Bulaevskii2012,Zotova2014}. Therefore, a saturation of the photon count rate as a function of $I$, observed in the best SSPDs~\cite{Baek2011,Marsili2011,Marsili2013,Kahl2015,Smirnov2018}, indicates that each absorbed photon is detected. This is the regime of unit internal efficiency, as verified both theoretically~\cite{Bulaevskii2012,vadim2,vodolazov} and in experiments with minimized coupling losses~\cite{Marsili2013,Kahl2015,Smirnov2018}.}

 Tuning our device in the regime of unit internal efficiency we are able to quantify the coupling efficiency $\eta_{couple}$. This experiment is designed as follows, \textcolor{black}{see the inset of Fig.~\ref{fig: thermalresistors}b for a schematics of our setup.} The SSPD is cooled to $T_b=1.7$\,K,  where it shows its best performance, and illuminated with 1.2\,pW CW optical radiation of wavelength 1.55\,$\mu$m through a single-mode fiber attached on top of the detector. {\bblue At a finite DC bias current,} the absorption of a single photon results in a voltage {\bblue pulse, further} amplified by broadband microwave room temperature amplifiers (50\,dB gain) and counted with a pulse counter Aligent 53131A. \textcolor{black}{The input optical power is measured at the entrance of the dipstick with the Ophir PD300-IRG calibrated power meter. $P_{in}$ was attenuated by the optical variable attenuator EXFO FVA-600. The coupling efficiency of the SSPD varies by $\sim20$\% depending on the polarization of the incident light and the experiment is performed at the maximum response. Note that the choice of the polarization is irrelevant for the measured $Z$}. {\bblue Fig.~\ref{fig: thermalresistors}b shows the bias current dependence of the measured detection efficiency, $D_E$, which is the ratio of the counted and input photon numbers. The observed saturation of $D_E$ at increasing current, see the dashed line in Fig.~\ref{fig: thermalresistors}b, is the evidence of the regime of unit internal efficiency. Here, the photon detection is limited only by the coupling efficiency and we obtain $\eta_{couple}=D_E\approx 52$\%, which is used to calculate $P_{abs}$ and $Z$ below.}

\subsection{\label{resistivity}Thermal resistance}

{\bblue The final step to obtain the thermal resistance, $Z=\delta T_e/P_{abs}$, is the measurement of the variation of the electron temperature in-phase with the modulated optical power. We track the small oscillation $\delta T_e$ by measuring the oscillating voltage on the SSPD biased with the DC current $I$ using a standard relation $\delta U = I(\partial R/\partial T_e)\delta T_e$. We expect that $T_e$ is the only parameter that controls the detector resistance in our experiment, therefore $\partial R/\partial T_e$ is the same as the derivative $dR/dT$ determined from the $T$-dependence of the resistance in equilibrium, when the bath and electron temperatures coincide, $T=T_e=T_b$.  Obviously, our approach is sensitive to $\delta T_e$ only in the vicinity of the resistive transition, where the $T$-dependence is strongest, hence the resolution is limited by the width of the transition. In order to expand the $T$ range available  
we use a perpendicular magnetic field which shifts the transition to lower temperatures and simultaneously smears it out.}

\begin{figure}[t]
\includegraphics[width=0.8\columnwidth,trim={0cm 7cm 3cm 1.2cm}, angle=0]{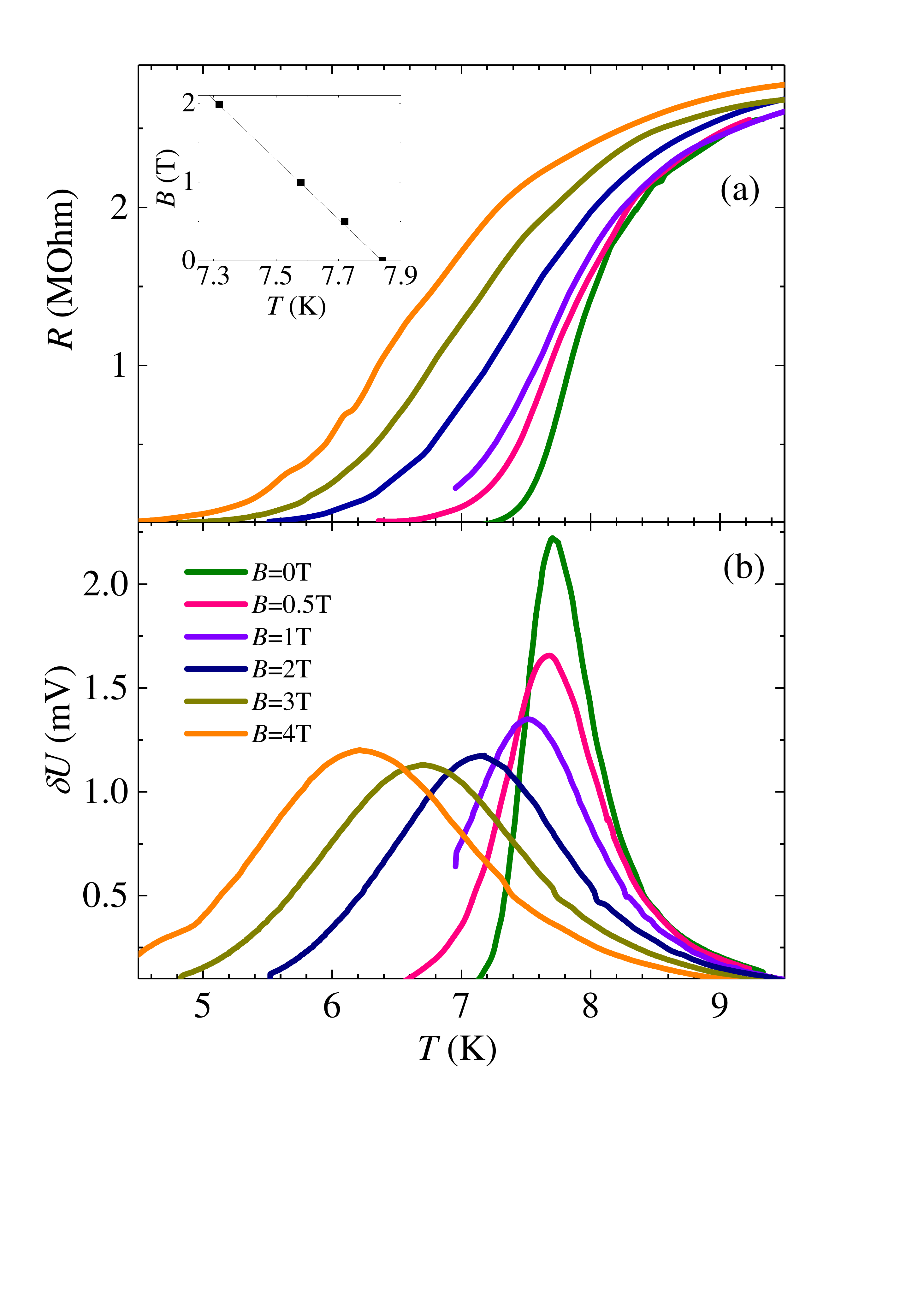}
 \caption{\bblue The $T$-dependencies of the sample resistance in equilibrium (a) and of the measured voltage response under optical radiation (b) for a set of magnetic fields (see the legend). In panel (b) the laser wavelength of 1.55 $\mu$m, the input laser power of $P_{in}=$320\,nW and the DC bias current of $I=100$\,nA were used. Inset: the $T$-dependence of the upper critical field  $B_{c2}(T)$ used to evaluate the density of states in NbN.  \textcolor{black}{The $T$-dependencies in panels (a) and (b) were taken by measuring the signals during a warm-up at discrete temperature values (15-20 points per curve). At each point the temperature was stabilized for at least 10 minutes with the accuracy of about $\pm$1\,mK.}}
 \label{fig: rt}
\end{figure}
 
{\bblue Fig.~\ref{fig: rt}a and Fig.~\ref{fig: rt}b show, respectively, the $T$-dependencies of the sample resistance in equilibrium and of the measured voltage modulation under the modulated optical radiation for a set of magnetic fields up to $B=$4\,T. The data of Fig.~\ref{fig: rt}a were smoothed and the derivative $dR/dT$ was obtained numerically. At increasing $B$ the $T_c$ decreases and the resistive transition smears out. The measured $\delta U$ changes accordingly. While the input optical power remains constant throughout the experiment, $\delta U$ follows the bell-shaped curve, which is slightly different from the calculated $dR/dT$ (not shown). As demonstrated below, for $B\geq1$\,T this difference is fully captured by the $T$-dependence of the thermal resistance $Z(T)$.
}

{\bblue The measured $\delta U$ in Fig.~\ref{fig: rt}b  translates in a temperature variation as small as $\delta T_e\sim10$\,mK. This value, averaged both in time and throughout the device, is much  smaller compared to the characteristic temperature variation $\delta T_e^{HS}$ within the HS formed after the photon absorption. Obviously, the validity of our approach is based on the assumption that the temperature dependencies of both $Z(T)$ and $R(T)$ can be approximated by linear on the scale of $\delta T_e^{HS}$. $\delta T_e^{HS}=\alpha E_{ph}/(C^n_eV)$, where $\alpha$ is the assumed fraction of the photon energy $E_{ph}=0.8$\,eV which goes into electron system, $C^n_e$ is the normal state thermal capacity of the electron system and $V$ is the HS volume. The HS volume is  $V=2r_{HS}wd$, where  $d=5.5$\,nm and $w=100$\,nm are, respectively, the thickness and width of the superconducting film and $r_{HS}=2(D\tau)^{1/2}\approx160\,{\rm nm}>w$ is the HS radius,  obtained from the experimental value of the diffusion coefficient $D$ and relaxation time of $\tau\sim50$\,ps typical for NbN. Based on the data of section~\ref{Bc2} and with $\alpha=0.4$, which correlates with our analysis below, we estimate $\delta T_e^{HS}\approx1.5$\,K. This value is indeed much stronger heating than the measured time- and device-averaged value of $\delta T_e\sim10$\,mK and comparable to the width of the resistive transition in a magnetic field. Still, the fact that the data in different $B$ coincide within the experimental uncertainty (Fig.~\ref{fig: z}) indicates that our assumption of the linear response is reasonable. Hence, the inevitable spatial and temporal inhomogeneity of the heating and the associated resistive response of the electron system is irrelevant for the purpose of $Z(T)$ measurement.}

\subsection{\bblue \label{Bc2} Density of states and electron heat-capacity }

{\bblue Transport measurements also contain the information on the $T$-dependence of the upper critical field 
$B_{c2}(T)$ in our sample, which is determined from the maxima of $dR/dT$ and plotted in the inset of Fig.~\ref{fig: rt}a.   }
As usual, from the linear part of this $T$-dependence we extract the diffusion constant in NbN $D=-({4k_B}/{\pi e})( {dB_{c2}}/{dT})^{-1}=0.35$ cm$^2$/s. {\bblue Next, using} the Sommerfeld model for free electron gas \cite{Kittel} we obtain a single-spin density of states at the Fermi-level $N(0)= (2\rho_n D e^2)^{-1}\approx20.3\,{\rm eV^{-1}nm^{-3}}$, {\bblue where $\rho_n$ is the normal state resistance  at $T=20$\,K, and evaluate the normal state thermal capacitance $C^n_e=2\pi k_B^2 N(0)T/3$, where, $k_B$ is the Boltzmann constant. We believe that this {\it ab-initio} estimate is well justified in our dirty NbN system, since the electron mean-free path is about 1$\rm \AA$ thereby the crystalline order is destroyed and a possible Fermi surface anisotropy is negligible.}

\section{\label{sec:results}Results}

{\bblue In Fig.~\ref{fig: z} we present  the temperature dependence of the measured thermal resistance $Z$ of our NbN SSPD for a set of magnetic field $B$ values, obtained as described in section~\ref{resistivity}. At each setting of $B$, see the legend, the data points cover a  temperature range corresponding to the width of a respective $\delta U$ peak  in Fig.~\ref{fig: rt}b. For different $B\geq1$\,T, the data ranges sufficiently overlap and demonstrate good consistency, indicating that our two-temperature model adequately describes the power response of the SSPD in finite magnetic fields. Altogether, this data covers the range between 5\,K and 9\,K and enables us to extract the power-law temperature dependence of the form $Z\propto T^{-3}$, see the dashed guide line in Fig.~\ref{fig: z}. 

\begin{figure}[t]
    \centering
\includegraphics[width=0.8\columnwidth, angle=0]{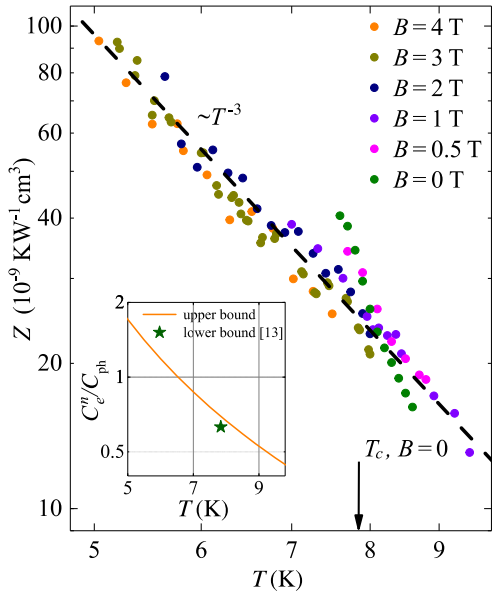}
\caption{Temperature dependence of the thermal resistance $Z$ for NbN SSPD at resistive transition. {\bblue The experimental datasets are obtained from the measured power-response of our SSPD  in different magnetic fields, as explained in section~\ref{resistivity}. The dashed guide line represents the power-law  dependence $Z\propto T^{-3}$ observed.  The $B=0$ critical temperature of the resistive transition is marked by an arrow. Inset: the ratio of the electron (normal-state) and  phonon thermal capacitances as a function of temperature. The solid line is the upper bound imposed by our experiment, as discussed in the text. The symbol is the lower bound, which follows from the theory of Ref.~\cite{vodolazov} for the parameters close to our SSPD detector.}}
 \label{fig: z}
\end{figure}

Notable in Fig.~\ref{fig: z} are sizeable deviations for the two lowest magnetic fields, for  which we find that the $T$-dependence becomes considerably stronger at $B\rightarrow0$. In this respect, our observation are reminiscent of the $B=0$ peculiarities of the relaxation time observed long ago in dirty Nb films~\cite{gershenzon1984} and more recently  in boron-doped diamond films~\cite{kardakovaPRB} by the AMAR technique. The origin of this puzzling behavior was associated~\cite{kardakovaPRB} with the longitudinal relaxation time in non-equilibrium superconductivity~\cite{tinkham}. This physics is certainly beyond the applicability of our two-temperature model, therefore the corresponding two datasets in Fig.~\ref{fig: z} are discarded in our analysis of the $T$-dependence of $Z$.}

{\bblue  As seen from the eq.~(\ref{eq:3}), in principle, the knowledge of $Z(T)$ allows to extract simultaneously both the electron-phonon and the phonon escape thermal resistances, provided their $T$-dependencies are sufficiently different. Unfortunately, this program is difficult to realize here. According to the literature, for thin NbN films $\tau_{eph}\propto T^{-1.6}$~\cite{gousev,gousev1994} and $\tau_{esc}\propto d$~\cite{Rall2010, henrich2013influence, cherednichenko}, with numerical pre-factors dependent on the film quality, substrate and, perhaps, other experimental details. Hence, assuming for the thermal capacities $C_e\propto T$ and $C_{ph}\propto T^3$, we expect the functional dependencies of $Z_{eph} \propto T^{-2.6}$ and $Z_{esc} \propto T^{-3}$, which are very close both to each other and to the experimental $Z(T)$ in Fig.~\ref{fig: z}.}

The uncertainty of the available data on the timescales $\tau_{eph},\tau_{esc}$~\cite{gousev, gousev1994, Rall2010, henrich2013influence, cherednichenko}, as well as the lack of the independent measurement of $C_e$ at the resistive transition, makes it unreliable to discriminate between the two contributions to $Z$ in eq.~(\ref{eq:3}). Instead, using the evaluated normal state electron thermal capacity ($C^n_e$, see the section~\ref{Bc2}) and the experimental $Z(T)$, we determine the upper boundary for the ratio $C^n_e/C_{ph}$. This is achieved via the relation~(\ref{eq:cecp}) in which the lower bound of the acoustic phonon escape time  in dirty NbN films is adopted from the Ref.~\cite{semenov1} $\tau^{low}_{esc}/d=8$\,ps/nm. Note, that the linear dependencies of relaxation time on $d$, established in Refs.~\cite{cherednichenko,henrich2013influence}, along with the typical time-scales $\tau_{esc}\gg d/s$, where $s$ is the sound velocity, imply extremely short mean-free paths for acoustic phonons in thin NbN films. Thereby the scenario of full-internal reflection of phonons proposed recently in WSi~\cite{Sidorova2018} is not applicable here. The result of our analysis is shown in the inset of Fig.~\ref{fig: z} by the solid line with the $T$-dependence of $C^n_e/C_{ph}\propto T^{-2}$ and the absolute value $C^n_e/C_{ph}\sim1$, which is in reasonable agreement with the estimates based on the Debye model in NbN, as discussed below. \textcolor{black}{Most intriguing is a comparison with the recent SSPD  theory~\cite{vodolazov}, which relates the $C^n_e/C_{ph}$ with the detection efficiency. We observe that  in the vicinity of the $B=0$ resistive transition the experimental upper boundary is close  to the theoretical lower boundary for SSPDs with the unit internal efficiency, see the  symbol in the inset of Fig.~\ref{fig: z}. This finding is also discussed below.}

\section{\label{sec:discussion}Discussion}

{\bblue The measurement of the absolute value of $Z(T)$ in Fig.~\ref{fig: z} is the central experimental result of our work. The data is consistent with the functional dependence as well as the order of magnitude estimate based on the available data on the relaxation time scales $\tau_{eph}$ and $\tau_{esc}$ in NbN~\cite{gousev,gousev1994,Rall2010, henrich2013influence, cherednichenko} and the estimates of $C_e$ and $C_{ph}$. This justifies our experimental approach, which has two clear advantages. First, the use of the optical frequencies ensures a uniform absorption efficiency by a potentially inhomogeneous superconducting film, in contrast, e.g., to a much simpler Joule heating approach~\cite{brown}. Second, the device and setup dependent radiation coupling efficiency can be accurately calibrated in the regime of a unit internal efficiency of our SSPD, providing the knowledge of the absolute absorbed optical power. The main drawback of our approach is the temperature resolution limited by the width of the resistive transition, which can be overcome by shifting the transition with the help of magnetic field, yet with certain reservations. In particular, it is not clear at present, whether the deviations of $Z(T)$ around $B=0$ from the general trend $Z\propto T^{-3}$ in Fig.~\ref{fig: z} capture the real changes in $Z$ or manifest a failure of the two-temperature model caused, e.g., by the non-equilibrium phenomena~\cite{kardakovaPRB} or by the vicinity of the BKT transition. 

The lower bound imposed on the phonon thermal capacity, $C_{ph}>\tau^{low}_{esc}/Z$, and translated in the upper bound of $C^n_e/C_{ph}$ in Fig.~\ref{fig: z} (inset, solid line) is also an intriguing result. Within the Debye model $C_{ph}=(12\pi^4/5)N_{i}k_B(T/\Theta_D)^3$ ~\cite{neilashcroft1987}, where $N_{i}$ is the ion density, we obtain an estimate of the Debye temperature $\Theta_D<430$\,K which reasonably compares with some reported values $\Theta_D=174-363$\,K \cite{Chockalingam174k,Geballe311k, Roedhammer330k, Geibel363k}. 

Luckily, more conclusive is a comparison with the recent ``photon-generated superconducting vortex model" of Ref.~\cite{vodolazov}. { In our SSPD operating in the unit internal efficiency regime at 1.7\,K the ratio of the measured superconducting critical current and the theoretical depairing current, $I_{dep}$, does not exceed 0.6. Hence, the saturation of detection efficiency in Fig.~\ref{fig: thermalresistors}b takes place about $I\approx0.4I_{dep}$. For NbN SSPD devices of similar width $w=128$\,nm, $T/T_c\approx0.2$ and $E_{ph}=0.8$\,eV the ``photon-generated superconducting vortex model" predicts the saturation  at $I/I_{dep}\approx0.5$ provided 
$\left(C_e^n/C_{ph})\right|_{T=T_c}\geq0.63$, see the upper dataset of Fig.~12 in Ref.~\cite{vodolazov} extrapolated to lower $T/T_c$. This lower bound, marked by the symbol in the inset of Fig.~\ref{fig: z}, is just about 15\% below the solid line, implying that the actual ratio $C^n_e/C_{ph}$ is quite close to the experimental upper bound. Thus, we speculate that, first, our SSPD resides on the boundary of the best materials for single-photon detection and, second, the thermal resistance between the NbN and the substrate is dominated by the phonon escape term, i.e. $Z\approx Z_{esc}\gg Z_{eph}$ in eq.~(\ref{eq:3}). \textcolor{black}{The former conjecture correlates with an empiric finding that  NbN-based SSPDs with a unit internal efficiency are trickier in fabrication as opposed to their analogues based on WSi, which have a much larger thermal capacities ratio~\cite{vodolazov}.} The latter one is less obvious and implies that $\tau_{esc}\gg\tau_{phe}$, i.e. the phonon escape time into the substrate is much longer than its life time limited by the electron-phonon interaction. Although it is difficult to draw a definitive conclusion based on present experiment, such a strong electron-phonon coupling regime might serve to explain the well known fact~\cite{Rall2010, henrich2013influence, cherednichenko,semenov1,Sidorova2018} that $\tau_{esc}$ in dirty superconducting films is more than an order of magnitude longer than the time of flight of a ballistic phonon in the direction of a substrate. }}

{\bblue In summary, we investigated the thermal properties of a NbN single-photon detector capable of unit internal detection efficiency in the infrared range. The key of our approach is the independent calibration of the apparatus' coupling losses, which provides the knowledge of the optical power absorbed by the NbN film. Using a resistive response of the device for thermometry in the vicinity of the resistive transition in zero and finite magnetic field we extract the thermal resistance $Z$ and analyze its temperature  dependence within the two-temperature model. Our results provide an estimate of the ratio of electron and phonon heat capacities and suggest that the thermal resistance is dominated by the process phonon escape from the NbN film into the substrate.}

\begin{acknowledgements}
We acknowledge discussions with D.Yu.~Vodolazov and E.S.~Tikhonov. The research is supported by the Russian Science Foundation project No. 17-72-30036. 
\end{acknowledgements}

\bibliographystyle{ieeetr}
\nocite{apsrev41Control}
\bibliography{main}

\end{document}